# A Content-based Centrality Metric for Collaborative Caching in Information-Centric Fogs

Junaid Ahmed Khan[‡], Cedric Westphal[*] and Yacine Ghamri-Doudane[†]

[‡]Univ Lyon, INSA Lyon, INRIA, CITI Lab, Villeurbanne, France
[*]University of California, Santa Cruz, CA, USA
& Huawei Technology, Santa Clara, CA, USA
[†]L3i Lab, University of La Rochelle, France
junaid.khan@inria.fr, cedric@soe.ucsc.edu, yacine-ghamri@univ-lr.fr

*Abstract*—Information-Centric Fog Computing enables a multitude of nodes near the end-users to provide storage, communication, and computing, rather than in the cloud. In a fog network, nodes connect with each other directly to get content locally whenever possible. As the topology of the network directly influences the nodes' connectivity, there has been some work to compute the graph centrality of each node within that network topology. The centrality is then used to distinguish nodes in the fog network, or to prioritize some nodes over others to participate in the caching fog. We argue that, for an Information-Centric Fog Computing approach, graph centrality is not an appropriate metric. Indeed, a node with low connectivity that caches a lot of content may provide a very valuable role in the network.

To capture this, we introduce a *content-based centrality (CBC) metric* which takes into account how well a node is connected to the *content* the network is delivering, rather than to the other nodes in the network. To illustrate the validity of considering content-based centrality, we use this new metric for a collaborative caching algorithm. We compare the performance of the proposed collaborative caching with typical centrality based, non-centrality based, and non-collaborative caching mechanisms. Our simulation implements CBC on three instances of large scale realistic network topology comprising $2,896$ nodes with three content replication levels. Results shows that CBC outperforms benchmark caching schemes and yields a roughly $3x$ improvement for the average cache hit rate.

*Index Terms*—Content Centric Networking, Distributed Content Caching, Fog Networking, Content Offload.

## I. INTRODUCTION

Information-Centric Fog Networking has been proposed to allow nodes and devices near the end-user edge to perform caching, communication and computation, thereby reducing the demand on the data center cloud and the communication over the Internet backbone. The goal of Information-Centric Fog Networking is to offload the content and some of its processing closer to its consumers in order to reduce cost, bandwidth consumption and network overhead.

The nodes around the end-user form a network from which the consumer can retrieve the content. However, this network is not as structured and organized as the managed operator network. To identify the structure of this fog network, it has been suggested to consider the graph centrality derived from the graph topology of the network.

Centrality [1], a concept from graph theory typically applied to social networks, is used to find important nodes in a graph. A high centrality score reflects a high topological connectivity for a node in the network. Typical centrality measures are: degree (the number of directly connected nodes), closeness (the average length of the shortest paths between the node and all other nodes in the graph); betweenness (the number of shortest paths between all pairs of nodes in the graph going through a specific node), and eigenvector centrality (a measure of node influence in the network).

Computing centrality on the topological graph yields interesting insights. Yet it fails to capture that, in an Information-centric network, the consumer is interested in connecting to the *content*, not to a specific *node*. To address this, we introduce the concept of Content-Based Centrality. We define a content-based betweenness centrality. Namely, for a node $v$, its CBC is calculated as the number of shortest paths from all the consumers to all the content which go through $v$. It is defined formally in Section III.

To show the usefulness of the CBC, we propose a collaborative caching method where a set of connected nodes can mutually self-organize into a fog for distributed content caching. We provide a CBC-based content placement algorithm, where popular content is cached at high CBC nodes, and in similar fashion nodes in the the fog cache content with decreasing CBC score and content popularity.

We evaluate our proposed CBC-based content placement by simulations with three realistic topologies and three content replication ratios. The results show that the proposed CBC-based content placement outperforms typical centrality schemes, or schemes without coordination within the fog. Our contributions can be summarized as follows:

- We introduce the concept of Content-Based Centrality (CBC), which we believe is better suited for Information-Centric Networks than traditional graph centrality;
- We introduce a scalable method to compute CBC without a priori knowledge of the content placement, based only upon replication rules;
- We present a CBC-based content placement algorithm which takes into account the number of paths to content passing through a node to find nodes as caches in the network;

- We evaluate our algorithm through simulations which show a significant improvement.

The remainder of the paper is organized as follows. Section II discusses the related work. In Section III we define our network model and introduce our content based centrality metrics. In Section IV, we discuss the performance evaluation and results. Section V concludes the paper along insights into future directions.

## II. RELATED WORK

Content Caching has been studied for some time by the research community, spanning a wide spectrum, including Small Cell Networks (SCNs) [2],[3], Content Distribution Networks (CDNs) [4] and Information/Content Centric Networking (ICNs/CCNs). For example, in [5], distributed cache management decisions are made in order to efficiently place replicas of information in dedicated storage devices attached to nodes in the network using ICN. Similarly [6] addresses the distribution of the cache capacity across routers under a constrained total storage budget for the network. The authors found that network topology and content popularity are two important factors that affect where exactly should cache capacity be placed. [7] looked at pushing content to the edge to anticipate network congestion, while [8] computed the capacity of an ad-hoc network of caches.

In a recent work [9], game theory is exploited for caching popular videos at small cell base stations (SBSs). Another work, [10] proposed a game theoretic approach in ICN to stimulate wireless access point owners to jointly lease their unused bandwidth and storage space to a content provider under partial coverage constraints. Both papers targeted a pricing model instead of providing an efficient content placement solution.

[11] defines a "conditional betweenness centrality" and uses this metric to chooses which nodes will cache the content. The authors in [12] proposed new social-aware metrics adaptable to dynamic topology in order to cache content at vehicles. Socially-Aware Caching Strategy (SACS) [13] for Content Centric Networks (CCNs) uses social information in order to privilege Influential users in the network by pro-actively caching the content they produce. The authors detect the influence of users within a social network by using the Eigenvector and PageRank centrality measures.

Another centrality based caching approach in CCN is presented in [14] where the sizing the content store is based upon centrality. The authors exploit different centralities (betweenness, closeness, stress, graph, eccentricity and degree) to heterogeneously allocate content store at nodes instead of homogeneous allocation. It is proposed that a simple degree centrality based allocation is sufficient to allocate content store. Similarly, [15] shows that a higher cache hit rate can be achieved if content is cached at high betweenness centrality nodes.

We argue that the topological connectivity only partially relates to the content, and there is a need to consider the content reachability in the network not addressed in the prior art. A well connected node in the topological graph is not necessarily closer to end-users requesting the content. To the best of our knowledge, we are the first to propose to compute a content-based centrality.

## III. CONTENT-BASED CENTRALITY

### A. System Model

*1) Connectivity Model:* We assume that nodes are connected to a fog network through local, ad hoc connections. The connectivity between nodes is modeled by a graph $G(V, E^v)$ where $V = \{v\}$ is the set of nodes and $E^v(t) = \{e_{jk}(t) \mid v_j, v_k \in V, j \neq k\}$ is the set of edges $e_{jk}(t)$ modeling the existence of a communication link between nodes $j$ and $k$. To characterize the concept of Fog network, the local network is represented by the undirected graph $G(L, E^l)$, the set of nodes $L = \{l\}$ represents different locations $l$ and the set of edges $E^l = \{e_{pq} \mid l_p, l_q \in L, p \neq q\}$ are the respective boundaries that connects different neighborhoods. To consider spatio-temporal content placement, we assume the time $T = (\overline{t_1}, \overline{t_2}, ...)$ as a sequence of regular time-slots, where the $k^{th}$ time-slot is represented as $\overline{t_k} = [t_k, t_{k+1})$. We assume that during a particular time-slot the content placement is relatively stable and the nodes will stay connected for a period of minutes or hours, depending on the application. Co-located or closely interconnected nodes can self-organize to form a fog $s \in V$ in a location $l$ for distributed content caching locally near end users, where $S = \{s\}$, $S \subseteq V$ is the set of fogs in all locations.

*2) Caching Model:* We define the set of known content as $X = \{x_1, ..., x_N\}$ for a catalog of $N$ pieces of content, where $x_j$ is an indivisible content chunk in the network. In the remaining of the paper, we will deal with individual content chunks $x \in X$, however a larger size content can be composed of several such content chunks. The nodes can fetch the content either from the service provider using the operator's infrastructure link or locally from peers in the fog, using a low-cost connectivity. The content popularity can be shared with the nodes using three approaches, (i) Offline method by the content operator as a control message. (ii) Local monitoring by the nodes taking into account the number of user interests for the content, and (iii) part of content header shared by the service provider.

### B. A Content-Based Centrality Metric

The Content-Based Centrality (CBC) is defined as the sum of the ratio of the number of shortest paths from all users to all content that passes through the node to the total number of shortest paths between all the (user,content) pairs. Figure 1 shows an example of CBC where the node $x_1$ is on the path from the users to the content in the server, thus, considered as a high CBC node. The other typical node-based centrality measures such as degree, closeness, betweenness and eigenvector consider the node $x_2$ as the highest centrality node, however, compared to $x_1$, placing content at $x_2$ do no achieve optimal caching. Mathematically, CBC can be represented as:

$$cbc(v) = \sum_{u \neq v \neq x} \frac{\sigma_{ux}(v)}{\sigma_{ux}}, \quad (1)$$

where $\sigma_{ux}(v)$ is the number of paths from user node $u$ to content $x$ that passes through the node $v$ and $\sigma_{ux}$ is the total number of paths from the user node $u$ to the content $x$. The CBC is then normalized by the number of possible paths to all content in the network such as:

$$normal(cbc(v)) = \frac{cbc(v) - \min_{w \in V}(cbc(w))}{\max_{w \in V}(cbc(w)) - \min_{w \in V}(cbc(w))}$$

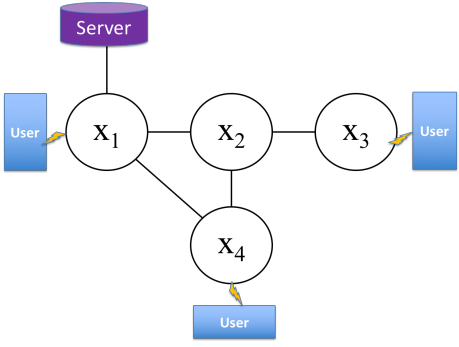

Figure 1: An example of Content-based Centrality

We use Betweenness centrality here, but we could use other form of centrality as well, such as Closeness or Eccentricity for instance.

*C. Scalable Computation of Content-based Centrality*

Computing the content-based centrality from Equation 1 is extremely difficult, since the catalog $X$ may be very large, and it requires knowing the content placement a priori, whereas we would like to compute the content-based centrality so as to guide the content placement.

Our solution is to note that the centrality of a piece of content does not depend on the specific chunk $x_j$, but rather, on the relative placement of the multiple copies of a chunk $x_j$. For instance, if $N_m$ objects are not cached anywhere in the fog network (and only at the origin server), their contribution to $cbc(v)$ in Equation 1 is equal to $N_m$ times the number of shortest paths to the origin server going through $v$ divided by the number of shortest paths to the origin server. We do not need to know which $N_m$ objects are cache misses, only that there are $N_m$ of them.

Further, we define by $\alpha$ the content replication factor that we allow at each node. $\alpha$ corresponds to our *replication policy*. For a total cache size of $b_v^t$ at the node, $\alpha b_v^t$ contains content (i.e. most popular content) common at all nodes, while $(1 - \alpha)b_v^t$ is the fraction of node caches with unique content compared to other nodes.

With the knowledge of $\alpha$, we can compute Equation 1 *without* knowing which content is placed at which node. However, we will later need to enforce this replication rule, by placing content such that a fraction $\alpha$ is replicated at all nodes, and the rest is unique to each node.

**Algorithm 1** Content Placement Algorithm
1: **INPUT:** $V, X, b_v^t, \forall v \in V, p_x, \forall x \in X, \alpha$
2: **OUTPUT:** $s \subseteq V, X_s, X_v \in X$
3: Initialize $s = \phi, b_{v'} = \phi, X_s = \phi$
4: **for** each location $l \in L$, time-slot $\bar{t} \in T$ **do**
5: $\quad v' = \arg\max_{v \in V}(cbc(v)), v' \notin s$
6: $\quad$ **while** $(b_{v'} \leq \alpha b_v^t)$ **do**
7: $\quad\quad b_{v'} \leftarrow \arg\max_{x \in X}(x, p_x),$
8: $\quad\quad$ Update $b_{v'}, X_{v'}$
9: $\quad\quad X_s = X_s \cup X_{v'}$
10: $\quad$ **end while**
11: $\quad$ **while** $(b_{v'} \leq b_v^t)$ **do**
12: $\quad\quad b_{v'} \leftarrow \arg\max_{x \in X, x \notin X_s}(x, p_x),$
13: $\quad\quad$ Update $b_{v'}, X_{v'}$
14: $\quad\quad X_s = X_s \cup X_{v'}$
15: $\quad$ **end while**
16: $\quad s = s \cup v'$
17: **end for**
18: **return** $s$

This specific replication policy is a simple first step, to demonstrate the feasability of a content-based centrality metric. We can define more sophisticated rules that allow to compute Equation 1 without knowing the specific content placement.

*D. Content Placement Algorithm*

The problem of content placement can be formulated as follows:

$$\begin{aligned} \underset{s}{\text{maximize}} \quad & \sum_{v \in s} X_v(l, \bar{t}), \forall l, \forall \bar{t} \\ \text{subject to} \quad & \sum_{x \in X} b_v^x(l, \bar{t}) \leq b_v^t(l, \bar{t}), \forall v, \end{aligned}$$

The objective function maximizes the content available at the nodes in the fog at each location and time-slot. The constraint consider the node buffer where an content cached at an individual node buffer should not exceed the maximum available threshold space. We also consider a defined replication factor $\alpha$ where for a node cache size $b_v$, $\alpha b_v$ contains content common at all nodes, while $(1-\alpha)b_v$ contains content different from other nodes.

The CBC is initially computed without any content placement, using the replication rule from the previous section. Then, the distributed content placement is optimized at co-located nodes using Algorithm 1. For a location $l$ and time-slot $\bar{t}$, the node $v' \in V$ with the highest CBC becomes the delegate to locally cache content in the fog as shown in Line 5. It initializes the fog formation by creating the subset $s \subset S$ and caches the most popular content, i.e content $x$ with maximum popularity $p_x$, while respecting the replication factor $\alpha$ for its respective storage buffer (Line 6). It continues caching content with decreasing popularity till the storage conditions based on the replication factor $\alpha$ are met. Once the storage buffer is at $\alpha b_v^t$, it updated the cached content and start filling the remaining content with decreasing popularity. The condition in Line 11 continues adding content till the node buffer $b_v^t$ attains its maximum threshold to cache content as indicated in our optimization constraint. The cached content is updated in the fog $s$ (Line 14) and the node adds itself to the fog (Line 16). Similarly, in a decreasing order of node CBC, the remaining nodes cache content in the fog $x$ with same content at $\alpha b_v^t$ and different at $(1-\alpha)b_v^t$. Thus, Algorithm 1 ensures a maximum amount of popular content are cached in the fog where the content set $X_v$ at each node is maximized which in turn maximizes our objective function.

## IV. NUMERICAL EVALUATION

We evaluate the proposed CBC along our collaborative caching algorithm using NS-3 where the named-data networking model of the ICN architecture is implemented. Three topologies are extracted from a realistic large scale trace of $2,986$ nodes (vehicles) in a $6X6km^2$ area, the city center of Cologne, Germany for one hour in order to validate the scalability of our caching approach. This allows us to evaluate the concept of fogs on realistic topologies reflecting connectivity in an urban environment. The topologies are the time snapshots of the network connectivity at 1 second granularity at the initial, at $30^{th}$ minutes and at $60^{th}$ minutes respectively.

### A. Simulation Scenario

The simulation scenario implements consumer nodes which generate interests for a pre-known content sequence of 100 unique items, each of 1024 kilobytes following a Zipf distribution (coefficient=1), i.e. frequent interests for more popular content. Any provider node already cached the content responds to the consumer interest. We allow intermediate nodes with uniform buffer size to perform in-network caching. We consider 30% of the nodes as consumers, 30% as providers (caching) nodes and the remaining nodes with disabled caching in order to accurately evaluate the performance of caching node. Each simulations is repeated 10 times with varying different nodes as consumers and producers. We also define $\alpha$ as a content replication factor where for a node buffer $b_v$, $\alpha b_v$ contains content which is same at all nodes in the fog, while $(1-\alpha)b_v$ is the buffer space for the cached content unique to each node.

As a first step, we randomly populate the caches of nodes in the fog with content in order to compute each node CBC. Random interests are generated. Then, we implement the collaborative placement Algorithm 1 where the highest CBC node caches the most popular content. Similarly content are populated at the remaining nodes' cache with decreasing content popularity and node CBC score. Besides CBC we implement three different caching approaches for comparison:

- **Centrality-based** Caching popular content at fog of high Degree, Closeness, Betweenness and Eigenvector centrality nodes. Algorithm 1 is implemented using each

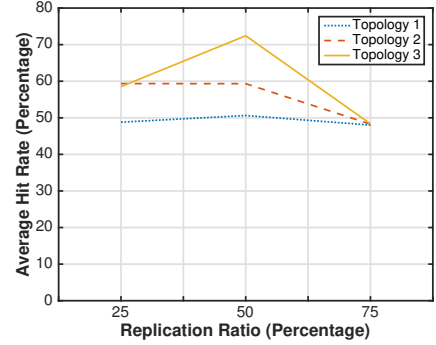

Figure 2: Average cache hit rate achieved by CBC on three different topologies with three content replication ratios.

  scheme where the most popular content is placed at the highest centrality node, then remaining content are placed at the nodes with decreasing node centrality and content popularity.
- **Non-centrality based** Social-unaware approach by caching greedily popular content at all nodes in the fog along a Least Recently Used (LRU) based content placement policy.
- **Non-collaborative based** approach where no fog is formed between nodes and each individual node cache indifferently. CBC is used to identify nodes.

The following performance metrics are used to evaluate CBC and the proposed collaborative caching algorithm:

- Cache Hits: It is the average number of content responds from the node cache, calculated as the ratio of the number of content responds to the number of received interests by the nodes.
- Success Rate: Different from the content responds, it is the average throughput of the nodes, calculated as the ratio of the number of interest responded with content or forwarded by the nodes to the generated interests.

### B. Simulation Results

*1) Cache Hits:* An important evaluation metric for content caching is the amount of cache hits. Therefore, we computed the cache hit rate for each scheme, (i) Fog with CBC, Degree, closeness, Betweenness and Eigenvector centrality, (ii) non-centrality based fog, and (iii) no fog where individual nodes cache indifferently. Figure 2 shows the average cache hit rate of the nodes in fog classified by CBC. We compare the hit rate of our approach for three topologies from the Cologne trace and three replication ratios ($\alpha = 25\%, 50\%$ and $75\%$). It is shown that the average cache hit rate achieved differs with respect to topology. Furthermore, we observe from the Topology 3 that between $\alpha = 25\%$, and $\alpha = 75\%$ replication ratio, there is an increase in the cache hit rate where around $\alpha = 50\%$, the maximum cache hit rate is achieved. It is also observed that there is a difference in the average hit rate between topologies. Thus, by keeping half of the content with similar content and the other half with different content is the optimal $\alpha$ which results in the maximum hit rate irrespective of the underlying topology.

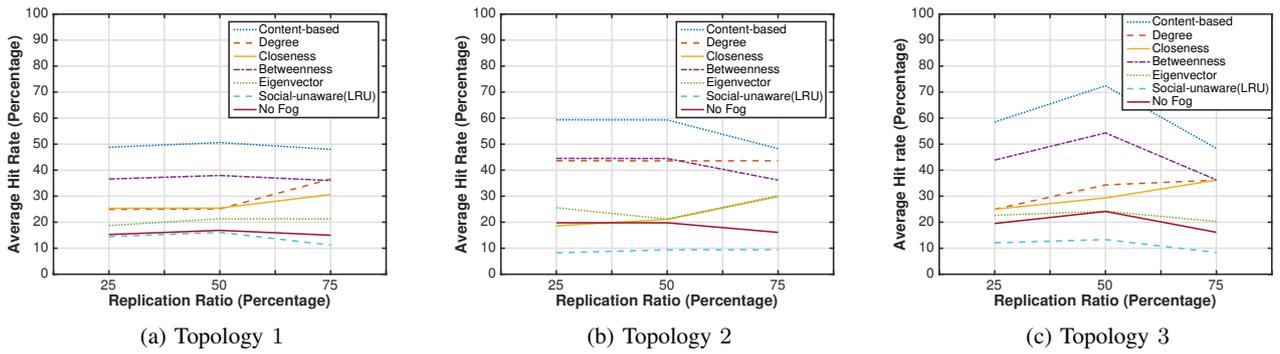

(a) Topology 1  (b) Topology 2  (c) Topology 3

Figure 3: Hit rate comparison for centrality-based (CBC, Degree, Closeness, Betweenness, Eigenvector), non-centrality based or social-unaware (LRU) and non-collaborative (no fog) based caching.

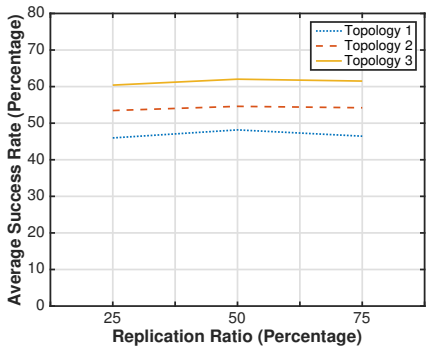

Figure 4: Average success rate achieved by CBC on three different topologies with three content replication ratios.

For a comparative analysis, we implement benchmark centrality schemes in the similar fashion by first placing popular content at the highest centrality node identified by each scheme. Then, following a decreasing node centrality order, we place the content at the set of nodes in the fog with decreasing content popularity order. Figure 3 shows the hit rate comparison where Figures 3a, 3b and 3c are the simulation results for each topology. It is clearly shown that irrespective of the network topology, the content-based centrality achieves a high cache hit rate when compared to all other approaches. It resulted in around $50\%$ hit rate for Topology 1, around $50-60\%$ for Topology 2 and up to $73\%$ for the Topology 3.

CBC is followed by betweenness centrality which achieved around $40\%$ average cache hit rate with a trend similar to CBC in Topology 1 and 3. All the centrality schemes resulted in substantially higher hit rate than the case without non centrality (social-unaware). For all the three topologies, this resulted in around $10\%$ hit rate with a maximum of $18\%$ cache hit rate for Topology 1.

Figure 3 also depicts the result by implementing a case where there is no collaboration between the caching nodes, thus, the case of no fog. It resulted in slightly better performance than the case of non-centrality based caching, though it achieves a lower average hit rate where it merely achieves around $20\%$ cache hit rate. Thus, the comparative analysis of cache hit rate on three different topologies and three content replication ratios reveals that content-based centrality achieves a better hit rate. It is because CBC the number of paths to content instead of nodes in a content-centric approach to classify nodes.

*2) Success Rate:* As an important network metric, we evaluate the success rate as the content responds or forwarding by the nodes in the Fog. In Figure 4, we show the average success rate achieved by the nodes in Fog classified using CBC. Similar to the cache hit rate, our analysis is based on three topologies and three replication ratios. Each topology resulted in a different success where Topology 3 shows the best performance with more than $60\%$ success rate for all replication ratios, followed by Topology 2 and Topology 1, thus suggesting that topology impacts performance. Another observation is that the success rate results from all topologies follow similar trend, thus validates the robustness of CBC when applied to different topologies. We see that by varying the replication ratio, a slight increase in the success rate is achieved for the $50\%$ replication ration where the success rate decreases towards $75\%$ replication ration. This suggest that for a node, it is optimal to keep half of the content similar to other nodes and the other half different from the other nodes.

Figure 5 shows a comparative analysis of the success rate achieved by different caching approaches. We observe that overall, CBC outperforms all other approaches, yielding an average success rate of around $50, 55$ and $60\%$ for Topology 1, 2 and 3 respectively. An interesting behavior is seen in the case of Topology 1 (Figure 5a) where fogs based on CBC, degree and betweenness centrality have similar success rate, though their score differs between replication ratio. For $25\%$ replication ratio, both degree and betweenness centrality show a higher success rate ($60\%$) than CBC, for $50\%$ replication ratio, CBC dominates betweenness centrality while degree stays higher, and for $75\%$ replication ratio, both degree and CBC show similar results (i.e. $47\%$ success rate), while betweenness dominates with around $58\%$ success rate. We investigated this behavior and found that since both betweenness and degree are node-centric metrics, and therefore better connected nodes at the relative center of the topology, they are acting as bridge between large number of nodes, thus resulting in a higher success rate. Nevertheless, nodes classified by CBC are equally better performing despite not being well placed in the network.

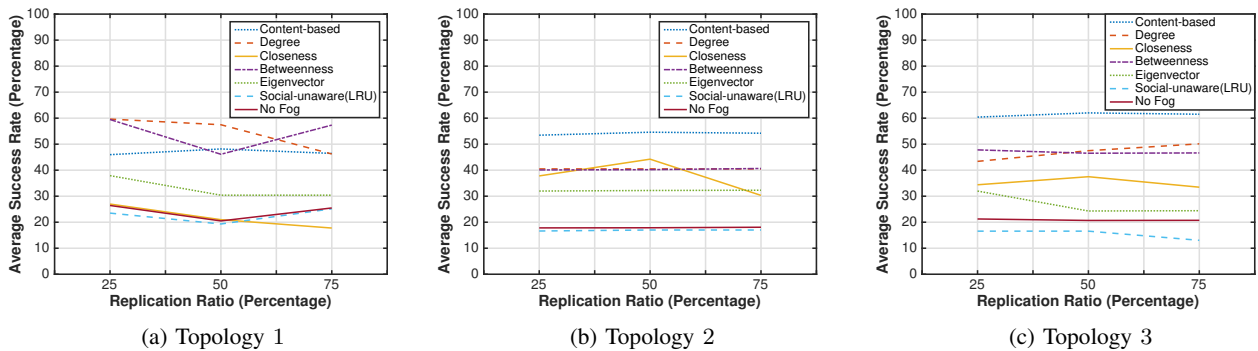

Figure 5: Success rate comparison for centrality-based (CBC, Degree, Closeness, Betweenness, Eigenvector), non-centrality based or social-unaware (LRU) and non-collaborative (no fog) based caching.

On the other hand, the non-centrality based socially-unaware resulted in the poorest performance in all topologies. Similarly, in the case without fog formation, i.e. non-collaboration resulted in poor performance with an overall success rate of less than 20%. A collaborative approach should definitely be used to for content caching.

## V. CONCLUSIONS AND FUTURE DIRECTIONS

Our results show that content placement in ICN can be efficiently managed by forming fogs allowing nodes to collaboratively cache at high content-based centrality nodes. "content" based centrality outperforms the node-centric approach to classify nodes for distributed caching.

It is challenging to find the appropriate nodes to be strategically selected for efficient content caching in the ICN network. In this paper, we targeted the problem by proposing a collaborative distributed caching approach at connected nodes near the network edge as in a fog network. To do so, we suggest to exploit graph properties to identify the suitable nodes for caching, however, unlike typical node-centric centrality scheme, we first presented CBC, a new Content-based centrality scheme where the number of paths to all content, instead of nodes, counts towards the node centrality. Then, we proposed an algorithm for collaborative content placement in the fog, where the nodes place popular content at high CBC nodes, followed by placing the remaining content with decreasing popularity at nodes with decreasing CBC score, according to a replication rule which allows to compute CBC without knowing the actual content placement.

We evaluated the benefits of CBC via simulations on three different topologies and varying three content replication levels. Results show that nodes in the fog based on CBC outperforms, in terms of cache hit rate and success rate, existing centralities, non-centrality based and non-collaborative caching approaches. Future work includes moving to dynamic topologies and study the impact of mobility on CBC. We plan to use CBC for content retrieval where the user interests are forwarded to high CBC nodes.